\documentclass{article}
\usepackage{arxiv}
\usepackage{indentfirst}

\usepackage{qcircuit}
\usepackage{braket}

\usepackage{titlesec}

\usepackage{array}
\usepackage{amsmath}
\usepackage{bm}
\usepackage{graphicx}
\usepackage{algorithm}
\usepackage{algpseudocode}
\usepackage{booktabs}
\usepackage{xhfill}
\usepackage[labelfont={bf,footnotesize}]{caption}
\usepackage{here}
\usepackage{parskip} 
\usepackage[utf8]{inputenc} 
\usepackage[T1]{fontenc}    
\usepackage{hyperref}       
\usepackage{url}            
\usepackage{booktabs}       
\usepackage{amsfonts}       
\usepackage{nicefrac}       
\usepackage{microtype}      
\usepackage{lipsum}
\usepackage{graphicx}
\usepackage{txfonts}
\graphicspath{ {./images/} }
\usepackage{multicol}
\usepackage{cite}
\title{Application of quantum computing to a linear non-Gaussian acyclic model for novel medical knowledge discovery}

\author{
 Hideaki Kawaguchi \\
  Quantum Computing Center \\
  Keio University\\
  3‑14‑1 Hiyoshi, Kohoku, Yokohama 223-8522, Japan \\
  \texttt{hikawaguchi@keio.jp} \\
  }

\begin{document}

\maketitle

\begin{abstract}
Recently, with the digitalization of medicine, the utilization of real-world medical data collected from clinical sites has been attracting attention. In this study, quantum computing was applied to a linear non-Gaussian acyclic model to discover causal relationships from real-world medical data alone. Specifically, the independence measure of DirectLiNGAM, a causal discovery algorithm, was calculated using the quantum kernel and its accuracy on real-world medical data was verified. When DirectLiNGAM with the quantum kernel (qLiNGAM) was applied to real-world medical data, a case was confirmed in which the causal structure could be correctly estimated when the amount of data was small, which was not possible with existing methods. Furthermore, qLiNGAM was implemented on real quantum hardware in an experiment using IBMQ. It is suggested that qLiNGAM may be able to discover new medical knowledge and contribute to the solution of medical problems, even when only a small amount of data is available.
\end{abstract}


\begin{multicols}{2}
\section*{Introduction}
The utilization of medical data, which is increasing with the digitalization of medicine, has been attracting attention ~\cite{1}. As a method of utilizing medical data so far, clinical trials such as randomized controlled trials have been conducted to establish scientific evidence. However, it has been reported that there are problems associated with clinical trials, including limited number of subjects owing to the strict selection/exclusion criteria, as well as time, cost, and ethical restrictions required for their implementation~\cite{2}. By contrast, real-world medical data, which are secondary data collected from clinical environments, have been attracting attention as data that are rapidly increasing with the digitalization of medicine, for example, disease registries, electronic medical record data, and claim data containing details of medical procedures~\cite{2,3}. Real-world medical data can be collected from a wide range of patients with lower cost, time, and ethical constraints than those of clinical trials.

Currently, the main method for analyzing real-world medical data is “causal inference,” in which the direction of a certain causal relationship is determined and then the causal effect, which represents the strength of the causal relationship, is estimated. As long as the direction of the causal relationship is known in advance, the methods for estimating causal effects have continued to evolve significantly in recent years with the improvement in statistical methods~\cite{4,5,6,7,8}. However, in the medical field where there are many uncertain cases, it is difficult to determine the direction of causality in advance in many situations. In these situations, “causal discovery,” which detects causal relationships from data, is important for the discovery of new knowledge~\cite{9}. Particularly in the case of real-world medical data, the number of variables to be handled is so large that it is not realistic to exhaustively confirm causal relationships in advance, and thus, the importance of causal discovery methods is expected to grow stronger.

The causal discovery algorithm is a method for identifying causal graphs that represent causal relationships among variables by determining the direction of causal relationships according to the data, with some assumptions. In the conventional method, there is a limitation in that the structure of the causal graph cannot be uniquely identified~\cite{10,11}. The linear non-Gaussian acyclic model (LiNGAM) is a causal discovery method that has recently attracted attention~\cite{12}. LiNGAM uniquely identifies the structure of causal graphs by making non-Gaussian assumptions about the data~\cite{12}. More precisely, it assumes that the error variables are non-Gaussian and independent, that each variable has a linear relationship, and that the causal graph is non-cyclic. LiNGAM is an algorithm that uses the independence assumed for the error variables to identify the parameters that determine the causal graph structure. Although LiNGAM has the advantage of being able to uniquely identify the causal structure, it has been pointed out that it cannot properly estimate the causal structure depending on the independence measure used to estimate the parameters~\cite{13}. Among the proposed independence measures, the method using the kernel method has been attracting attention~\cite{13}. It is known that the kernel method can determine independence with high accuracy and can estimate the causal structure with high accuracy even when the error variable, which is assumed to be non-Gaussian, is actually close to a Gaussian distribution or when there are outliers in the data~\cite{13}. In the literature~\cite{13}, independence measures were designed using Gaussian kernels, but an improvement is to seek kernels that can construct independence measures with even better accuracy.

Recently, quantum kernels, which apply quantum computing to kernel methods, have been attracting attention~\cite{14,15,16}. In conventional (or classical) computers, the state of a bit is either 0 or 1; however, quantum computers use quantum bits, or qubits, which can take superpositions of 0 and 1, and quantum mechanical principles such as quantum entanglement for information processing. Because qubits are sensitive to noise and the superposition state is broken after a certain period of time, a fault-tolerant quantum computer (FTQC) that can correct errors caused by noise is required; however, there are still many technical and essential problems to be solved for realizing FTQC. By contrast, a quantum computer called noisy intermediate-scale quantum (NISQ) computer, which does not have an error correction function, has been realized in the last few years~\cite{17,18}. For an experiment on quantum kernels using a NISQ computer, the work of applying quantum kernels to support vector machine frameworks is known~\cite{15}. Kernel methods embed data in a high-dimensional feature space to facilitate analysis, whereas quantum kernels use quantum circuits to construct the kernel. With the use of the superposition state and of the information in the high-dimensional feature space, it is expected that quantum computers can efficiently construct functions that are difficult to represent with conventional computers.

Quantum computing and kernel methods can both be viewed as efficient ways to perform computations in the Hilbert space~\cite{14}. This commonality is the focus of this study, and it is proposed that the causal discovery algorithm can be improved by using quantum kernels for the independence measure in the LiNGAM using kernel methods. To the best of my knowledge, this is the first study to apply quantum computing to LiNGAM. The causal discovery algorithm using quantum kernels is applied to real-world medical data and its performance is examined to determine whether it can contribute to the detection of new medical knowledge. This study shows the potential usefulness of utilizing quantum computing as a way to improve the performance of LiNGAM.
\section*{Method}
\subsection*{Causal discovery algorithm}
\subsubsection*{DirectLiNGAM}
There are two main types of LiNGAM: ICA-LiNGAM~\cite{12}, which uses independent component analysis, and DirectLiNGAM~\cite{19}, which is based on regression analysis. On the basis of the findings of existing literature~\cite{13}, DirectLiNGAM was more accurate for LiNGAM using a Gaussian kernel, and therefore, DirectLiNGAM was used as the basis for this study.

In DirectLiNGAM with a total of \emph{p} variables, a single regression
analysis is performed \emph{p}-1 times with each variable except a fixed
\emph{j}-th variable as an explanatory variable. Here, because it is
known that an explanatory variable and a residual become independent
when a single regression analysis in the correct direction is
conducted~\cite{19}, the explanatory variable at the time when the
independence measure between the explanatory variable and the residual
becomes the lowest is set as the parent node of the objective variable.
Then, the same operation is performed on the other variables, except for
the explanatory variable, and the arrows between the variables are
sequentially determined. For more information on DirectLiNGAM, see refs.
~\cite{13,19}.

\subsubsection*{Independence measure}
Improving the accuracy of DirectLiNGAM requires setting an appropriate
independence measure. For the independence measure for DirectLiNGAM, the
independence measure with a normalized cross-covariance operator
(NOCCO)~\cite{20}, which is independent of the shape of the kernel, was
selected for a more appropriate application of the quantum kernel. More
specifically, the fact that the Hilbert-Schmidt norm of the NOCCO
approaches zero was used as a condition for a pair of random variables
$(X, Y)$ to be independent
(i.e., $X \Perp Y$).
\begin{equation} \label{equation1}
    I^{N O C C O}(X, Y)=\left\|V_{Y X}\right\|_{H S}^{2}
\end{equation}
Here, $V_{Y X}$ is NOCCO of $X$ and $Y$, and $\|\cdot\|_{H S}$ represents the Hilbert-Schmidt norm. If the variables $X$ and $Y$ are independent, $I^{N O C C O}(X, Y)$ in equation (\ref{equation1}) will be zero. Furthermore, given a finite sample $(X_1,Y_1),\ldots,(X_n,Y_n)$, the empirical independence measure $\hat{I}_{n}^{N O C C O}(X, Y)$ can be calculated as an estimator, as shown in equations (\ref{equation2}) and (\ref{equation3}) below. 
\begin{equation}\label{equation2}
    \hat{I}_{n}^{N O C C O}(X, Y)=T r\left[R_{Y} R_{X}\right]
\end{equation}
\begin{equation}\label{equation3}
    R_{Y}=G_{Y}\left(G_{Y}+n \varepsilon_{n} I_{n}\right)^{-1}, R_{X}=G_{X}\left(G_{X}+n \varepsilon_{n} I_{n}\right)^{-1}
\end{equation}
Here, $G_X$ and $G_Y$ are the centralized Gram matrices of each random variable and $\varepsilon_{n}$ is the regularization constant. For a more detailed explanation of the independence measure with NOCCO, including the introduction of equation (\ref{equation2}), see ref.~\cite{20}.

\subsection*{Quantum computing and kernel method}
\begin{figure}[H]
\centering  
\includegraphics[clip, width=\linewidth]{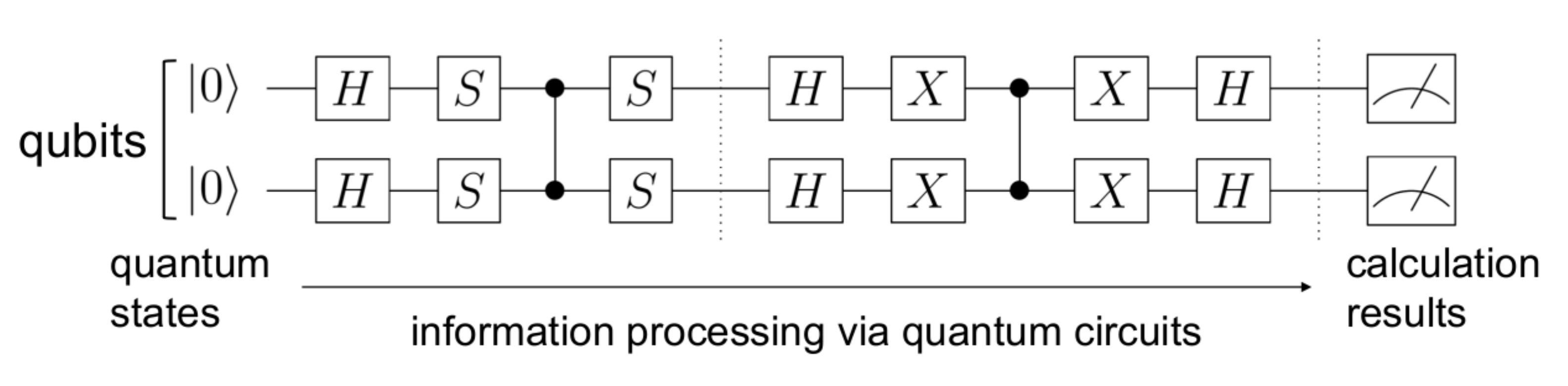}
\caption{\footnotesize\textbf{An example of quantum circuits.} This shows a circuit that performs a quantum calculation called Grover's algorithm. In quantum computation, qubits representing $|0\rangle^{n}$ as the initial states are prepared and passed through a quantum circuit to obtain the results. The wires mean that the quantum states are passed through as they are, and each gate marked with an alphabet changes the quantum state. The rightmost blocks with pictures of meters mean the observation of the quantum state.}    
\label{figure1}
\end{figure}

Recently, quantum computing has been attracting attention as an application technology for machine learning systems. Quantum computing uses qubits, which are superpositions of 0 and 1, and quantum circuits to process information. Figure\ref{figure1} shows an example of a quantum circuit in a gate-based quantum computing framework, in which the quantum state represented by a qubit is passed through a quantum circuit to perform information processing.

Quantum kernel is a typical technique of machine learning algorithms for NISQ computing, which are hybrid algorithms for quantum computing and conventional computing. The kernel method uses a feature map $\Phi$ to transform a given data ${{x}_{i}}$ taken from the original space $\mathcal{X}$ to a higher-dimensional Hilbert space $\mathcal{H}$.
\begin{equation} \label{equation4}
\Phi: \mathcal{X} \rightarrow \mathcal{H}, {x}_{i} \rightarrow \Phi\left({x}_{i}\right)
\end{equation}
\hspace*{1em}
The kernel method takes advantage of the fact that feature data mapping can be achieved without worrying about the dimensions of the feature space $\mathcal{H}$. The basic idea of the kernel method is that it is not necessary to know the format of the feature map explicitly; instead, the overlap of points in the feature space is simply taken as a measure of similarity between the features of any two data samples, which is called the kernel function. More specifically, the similarity between data $({x}_{i},{x}_{j})$ in the feature space is expressed using the inner product $\langle\cdot, \cdot\rangle$ and the kernel function $k(\cdot,\cdot)$ as shown in equation (\ref{equation5}) below.
\begin{equation} \label{equation5}
k({x}_{i}, {x}_{j})=\left\langle\Phi({x}_{i}), \Phi(x_{j})\right\rangle
\end{equation}
The kernel function can also be expressed as a matrix in the form of a Gram matrix as follows:
\begin{equation} \label{equation6}
K_{i j}=k({x}_{i}, {x}_{j})
\end{equation} 

Kernel methods embed data in a high-dimensional feature space to facilitate analysis, whereas quantum kernels use quantum circuits to construct the kernel. Quantum computers and kernel methods are very similar in principle in that they map information into a large space, but without the need for explicit computation in doing so. In the kernel method, access to the feature space is performed by the inner product of the kernels and feature vectors, whereas in the quantum computer, access to the Hilbert space of the quantum states is expressed by the inner product of quantum states. In the next section, the quantum kernels used in this study are discussed.
\subsection*{Quantum kernel estimation}
\begin{figure}[H]
\centering  
\includegraphics[clip,scale=0.9]{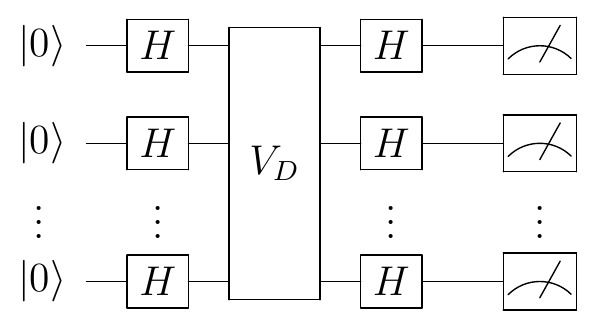}
\caption{\footnotesize\textbf{IQP circuit.} First, the gates labeled $H$ (Hadamard gate) are applied to the initial states of $|0\rangle^{n}$ to create the $|+\rangle^{n}$ state. Next, a gate $(V_D)$ consisting of diagonal matrices consisting of a high number of polynomials is applied. Finally, the $X$ measurement is performed by applying again the gates labeled $H$ and then performing the observation.}    
\label{figure2}
\end{figure}

First, the kernel function of the feature mapping is related to the inner product of the quantum states and equation (\ref{equation6}) is expressed in terms of the quantum kernel as follows:
\begin{equation} \label{equation7}
K_{i j}=\left|\left\langle\Phi(x_{j}) \mid \Phi(x_{i})\right\rangle\right|^{2}
\end{equation}
where $\Phi(\cdot)$ represents a quantum feature mapping and $\langle\cdot\mid\cdot\rangle$ represents the inner product of two quantum states in the Hilbert space. For the actual calculation of equation (\ref{equation7}) on a quantum computer, a quantum state $\Phi({x})$ needs to be created by acting a quantum gate $U({x})$, which represents some unitary transformation, on the initial state $|0\rangle^{n}$ of the quantum circuit, and the quantum state is created as shown in the following equation:
\begin{equation} \label{equation8}
U({x})|0\rangle^{n}=|\Phi({x})\rangle
\end{equation}
\hspace*{1em}
One of the motivations for using quantum kernels in LiNGAM is to search for quantum kernels that are difficult to reproduce using a classical computer. To construct $U({x})$ with a quantum circuit, an instantaneous quantum polynomial time (IQP) circuit was selected, which is known to have difficulty in accurately estimating the output probability distribution using a classical computer~\cite{21}. The IQP circuit is a circuit in which $|0\rangle^{n}$ as the initial states are multiplied by the Hadamard gates to create $|+\rangle^{n}$, which are then applied to the quantum gates consisting of only a few polynomial diagonal matrices, and finally, the ${X}$ measurement is performed (Fig. \ref{figure2}). 
If the IQP circuit is represented in mathematical terms, the layers that apply the Hadamard gates $H^{\otimes n}$ and that can be represented by a polynomial number of diagonal matrices $V_D({x})$ are repeated, as shown below.
\begin{equation} \label{equation9}
U_{I Q P}(x)=H^{\otimes n} V_{D}(x) H^{\otimes n} \cdots V_{1}(x) H^{\otimes n}
\end{equation}

In this study, the quantum gates $V_D({x})$, which consist of a polynomial number of diagonal matrices, were constructed in two layers: for one qubit, the $U$1 gate was applied to all qubits, and then for two qubits, the controlled-$U$1 ($CU$1) gate was linearly adapted to each neighboring qubit. The $U$1 and $CU$1 gates can be respectively expressed in matrix form as follows:

\begin{equation} \label{equation10}
U 1(\lambda)=\left(\begin{array}{cc}
1 & 0 \\
0 & e^{i \lambda}
\end{array}\right)
\end{equation}

\footnotesize
\begin{equation} \label{equation11}
C U 1(\lambda)=|0\rangle\langle 0|\otimes I+| 1\rangle\langle 1| \otimes U 1(\lambda)=\left(\begin{array}{cccc}
1 & 0 & 0 & 0 \\
0 & 1 & 0 & 0 \\
0 & 0 & 1 & 0 \\
0 & 0 & 0 & e^{i \lambda}
\end{array}\right)
\end{equation}
\normalsize
Here, the real data via the feature map function can be substituted for $\lambda$ in equations (\ref{equation10}) and (\ref{equation11}) and implemented in the quantum circuit.

\begin{figure}[H]
\centering  
\includegraphics[clip,scale=1.0]{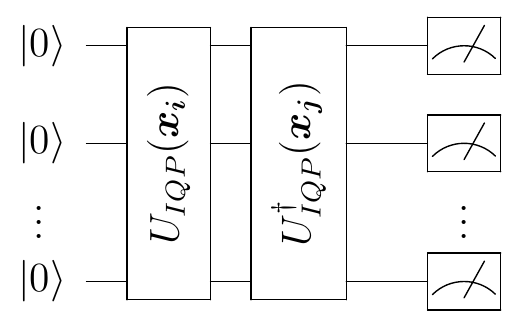}
\caption{\footnotesize\textbf{Inversion test.} The circuit is divided into the first half and the second half, and the dagger of the first half is taken. $U_{IQP}({x})$ represents equation (\ref{equation9}), which consists of layers with Hadamard gates $H^{\otimes n}$ and a polynomial number of diagonal matrices $V_D({x})$.}    
\label{figure3}
\end{figure}

For the quantum kernel estimation with a NISQ computer, there are two typical methods for quantum circuits: the swap test~\cite{22} and the inversion test~\cite{15}. In this study, an inversion test that required fewer gates was performed. To represent equation (\ref{equation7}) in a quantum circuit, in the inversion test, the circuit was divided into the first half and the second half and the dagger of the first half was taken and applied to the circuit in the second half (Fig. \ref{figure3}). With the use of equations (\ref{equation7})-(\ref{equation9}), the equation to be obtained can be expressed as follows:
\begin{equation} \label{equation12}
K_{i j}=\left|\left\langle 0^{n}\left|U_{I Q P}^{\dagger}({x}_{j}) U_{I Q P}({x}_{i})\right| 0^{n}\right\rangle\right|^{2}
\end{equation}
Herein, quantum DirectLiNGAM (hereinafter referred to as qLiNGAM) is proposed, which utilizes a quantum kernel as an independence measure for DirectLiNGAM. qLiNGAM is organized as follows: equation (\ref{equation12}) is used to calculate the Gram matrix in equation (\ref{equation3}) of the $\hat{I}_{n}^{N O C C O}$ calculation. The pseudocode for the algorithm is as follows:

\begin{tabular}{l} \toprule
\hspace*{-0.5em}\textbf{qLiNGAM} \\ \midrule
1) \textbf{Input:}\\
\hspace*{1.0em} $p \times n$ data matrix ${X}$,\\
\hspace*{1.0em} set $U$ of the subscripts of all ${x}_{i}\in {X}$,\\
\hspace*{1.0em} initialized ordering list of variables ${K}=\phi$,\\
\hspace*{1.0em} $m:=1$.\\
2) Repeat until $p$-1 subscripts are added to ${K}$.\\
\hspace*{1.0em} a) Regress $x_i$ on $x_j$ for all $i\in U-K(i \neq j)$ and \\ 
\hspace*{1.0em} derive the residual data matrix ${R}_{j}$ from the data\\
\hspace*{1.0em} matrix ${X}$ for all $j\in U-K$.\\
\hspace*{1.0em} b) Find a top node variable $x_t$ using the\\ 
\hspace*{1.0em} independence measure in equations (\ref{equation2}) and (\ref{equation3}):\\
\hspace*{3.0em}$x_{t}=\arg \min _{j \in U-K} T\left(x_{j} ; {U}-{K}\right)$,\\
\hspace*{3.0em}$T\left(x_{j} ; U\right)=\sum_{i \in U, i \neq j} \hat{I}_{n}^{N O C C O}\left(x_{j}, r_{i}^{(j)}\right)$,\\
\hspace*{1.0em} where the Gram matrices in $\hat{I}_{n}^{N O C C O}\left(x_{j}, r_{i}^{(j)}\right)$ are\\
\hspace*{1.0em} calculated using the quantum circuits in\\ 
\hspace*{1.0em} equation (\ref{equation12}).\\
\hspace*{1.0em} c) Add the subscript $t$ of the variable $x_t$ to $K$\\
\hspace*{1.0em} d) Let ${X}:={R}_{({t})}$ and $m:=m+1$\\
3) Add the subscript of the remaining variable to $K$.\\
4) Construct the strictly lower triangular structure\\
of the connection matrix $B$ using $K$.\\
5) Estimate the connection strength $b_{ij}$, which is\\
an element of $B$ using the data matrix $X$.
\\ \bottomrule
\end{tabular}

\section*{Results}
\subsection*{Preliminary experiment settings and model tuning}

\begin{figure}[H]
\centering 
\includegraphics[clip,scale=0.3]{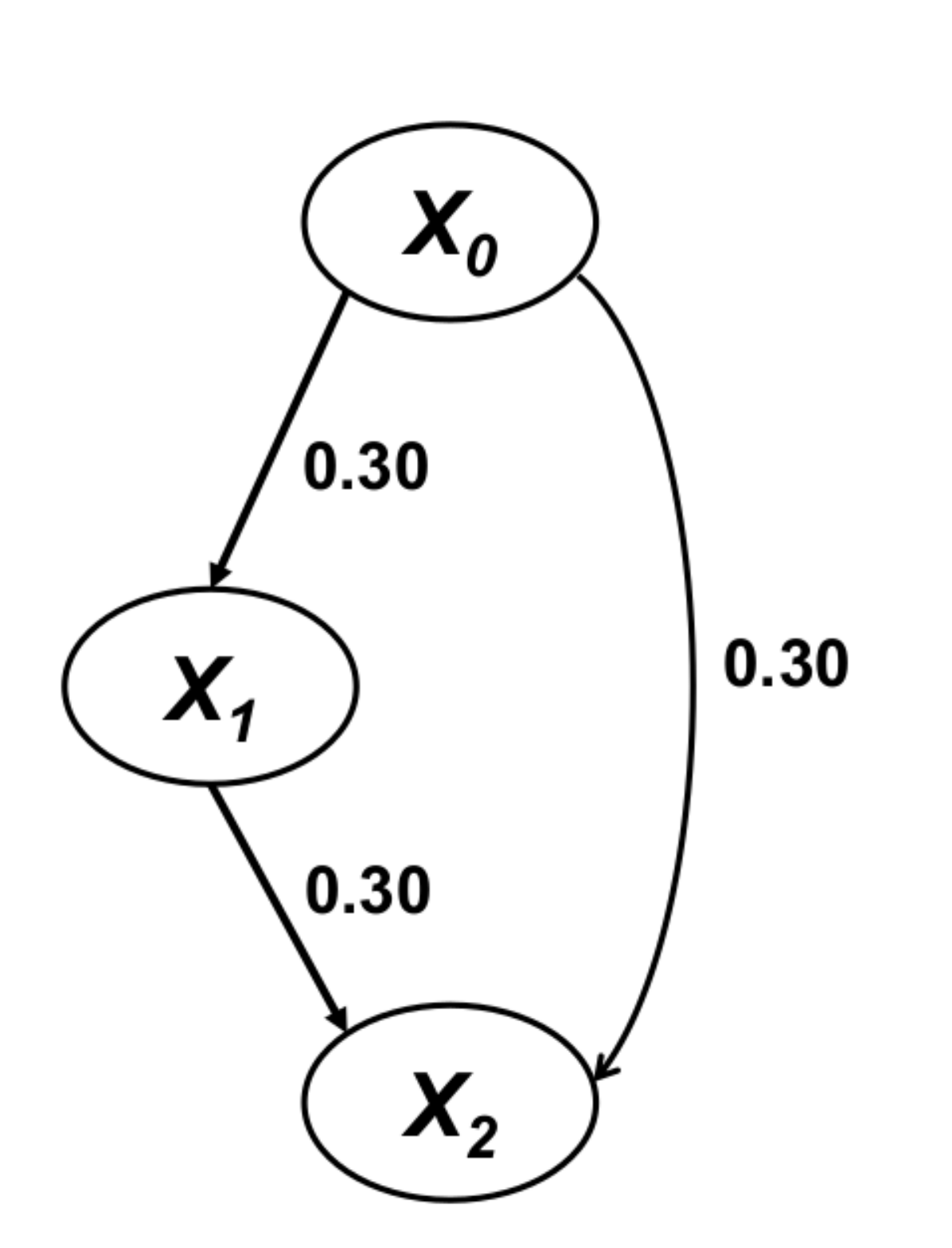}
\caption{\footnotesize\textbf{Causal relationship based on the artificial datasets.} The ellipses represent variables, and the arrows represent the directions of causality. The numbers attached to the arrows are the causal effect, which represent the coefficients in equation (\ref{equation13}).}    
\label{figure4}
\end{figure}

First, a simulation using artificial data was conducted. With a sample size of 100 and 3 variables, a dataset with an error term $e$ generated from the Laplace distribution ($\mu= 0, \lambda=1$) was created. The relationship between the three variables is as shown in equation (\ref{equation13}), which is represented in Fig. \ref{figure4}.
\begin{equation}\label{equation13}
\begin{aligned}
&X_{0}=e \\
&X_{1}=0.3 \times X_{0}+e \\
&X_{2}=0.3 \times X_{1}+0.3 \times X_{0}+e \\
&e \sim \frac{1}{2 \lambda} \exp \left(-\frac{|x-\mu|}{\lambda}\right)
\end{aligned}
\end{equation}

A total of 100 data sets with the structure of equation (\ref{equation13}) were prepared by changing the random seed and the error term. qLiNGAM was applied to 100 artificial datasets to determine the number of datasets that could identify the same causal structure as in Fig. \ref{figure4}. qLiNGAM and DirectLiNGAM with a Gaussian kernel, a typical conventional kernel, were applied to each dataset, and the number of correct causal structures identified was compared.

In constructing an IQP circuit, it is important to select a feature map function that corresponds to $\lambda$ in equations (\ref{equation10}) and (\ref{equation11}). In this study, the activation functions that are widely used in the field of neural networks were selected. Although sigmoid, tanh, and rectified linear unit are well-known activation functions, the ${tanh} \mathchar`-{shrink}$ function was selected, which is an activation function with continuity, as shown in equation (\ref{equation14}), for use as the phase of the quantum gates. In addition, when the data were inputted into the feature map function, they were normalized to mean 0 and variance 1, and then doubled for data scaling.
\begin{equation}\label{equation14}
{tanh} \mathchar`-{shrink}(x)=x-{tanh} (x)
\end{equation}

Based on the IQP circuit, the number of qubits to be used in the circuit and the number of layers of $V_D $ in equation (\ref{equation9}) to be repeated (called depth) are important variables in the construction of the quantum kernel. In this study, the number of qubits and number of depths were set to 5 and 2, respectively, which presented the best results using 10 data samples created using a random seed that were different from the aforementioned artificial data. For the implementation of the quantum kernel used in qLiNGAM, a quantum circuit was built with the Python library cirq~\cite{23} and quantum calculations were performed using qulacs~\cite{24}.

Table \ref{Table1} shows the experimental results. Both qLiNGAM and DirectLiNGAM with Gaussian kernels were able to identify the correct causal structure shown in Fig. \ref{figure4} for 39 out of 100 artificial datasets, whereas for 38 datasets, neither was able to identify it. In 14 datasets, only qLiNGAM was able to identify the correct causal structure, and in 9 datasets, only DirectLiNGAM with Gaussian kernels was able to identify it. Thus, qLiNGAM was able to estimate 14 datasets of the causal structures that DirectLiNGAM with Gaussian kernels was unable to detect.
\begin{table}[H]
\caption{\footnotesize\textbf{Results of the experiment using the artificial data} }    
\label{Table1}
\scalebox{0.68}{\begin{tabular}{p{3.2cm}cccc}\toprule
& Both & Neither & \begin{tabular}{c} Only\\qLiNGAM \end{tabular} & \begin{tabular}{c} Only\\DirectLiNGAM\end{tabular} \\ \midrule
\begin{tabular}{l}Number of \\correct causal \\structures identified \end{tabular}&\large39&\large38&\large14&\large9\\ \bottomrule
\end{tabular}}
\end{table}
\normalsize

\subsection*{Experiments with real-world medical data: Part 1}
Next, qLiNGAM was applied to real-world medical data. The first dataset used was the UCI Heart Disease Data Set~\cite{25}, which is an open-source dataset in the field of cardiology. Two datasets were prepared in advance: a full dataset of 297 cases from which records containing missing values were deleted, and a short version of the dataset from which 100 cases were randomly selected. The circuit configuration and associated parameters were exactly the same as those of qLiNGAM designed for the artificial data, and cirq and qulacs were used for the implementation.

From the UCI Heart Disease Data Set, three variables were used: ‘age,’ a variable representing age; ‘cp,’ a variable representing the four types of chest pain (1: typical angina; 2: atypical angina; 3: non-anginal pain; 4: asymptomatic), and ‘exang,’ a variable representing the presence or absence of exercise-induced angina (1: yes; 0: no). Of these, ‘exang’ is a factor that contributes to the classification of ‘cp’; therefore, a causal structure with an arrow drawn from ‘exang’ to ‘cp’ seems to be clinically valid.

\begin{figure}[H]
\centering  
\includegraphics[clip,scale=0.3]{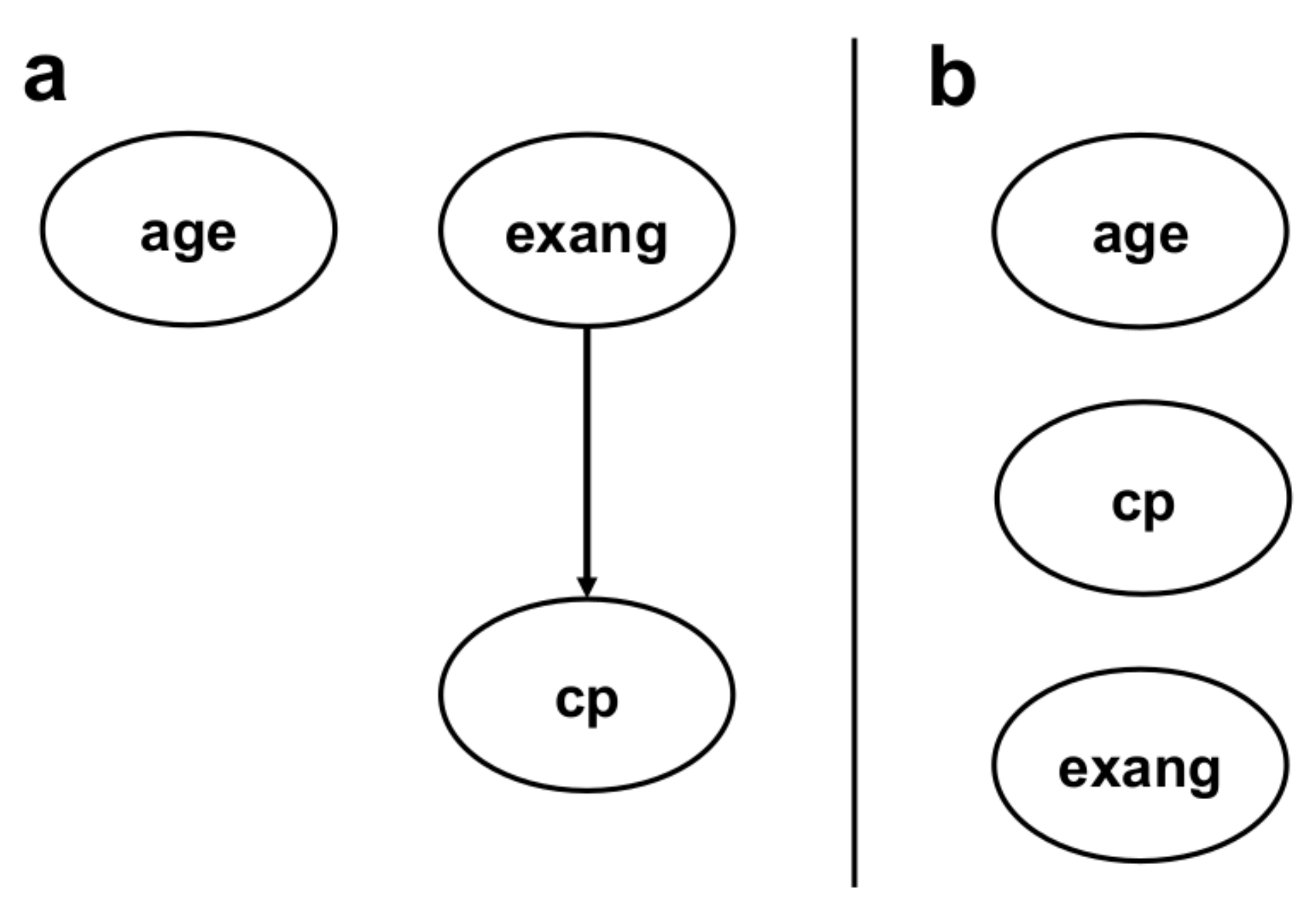}
\caption{\footnotesize\textbf{Experimental results using the UCI Heart Disease Data Set.} The ellipses represent the variables, and the arrows represent the directions of causality. \textbf{a}, A clinically valid causal structure, because of an arrow drawn from ‘exang’ to ‘cp’. \textbf{b}, Not a clinically valid causal structure in which all variables are independent. ‘age’, a variable representing age; ‘cp’, a variable representing the four types of chest pain; ‘exang’, a variable representing the presence or absence of exercise-induced angina.}    
\label{figure5}
\end{figure}

The experiment results showed that qLiNGAM identified the causal relationship described in Fig. \ref{figure5}a whether using the full version of the data with all 297 cases or the short version with only 100 cases. In this relationship, an arrow is drawn from ‘exang’ to ‘cp,’ which means that a clinically valid relationship was detected. By contrast, DirectLiNGAM with Gaussian kernels identified the causal relationship in Fig. \ref{figure5}a when using the full version of the data with all 297 cases and the one in Fig. \ref{figure5}b when using the short version with only 100 cases. The relationship in Fig. \ref{figure5}b is not a clinically valid result, where all variables are independent. These experimental results confirmed the case where the qLiNGAM was able to identify clinically valid causal relationships even with a smaller amount of data.

\subsection*{Experiments with real-world medical data: Part 2}
The next real-world medical data source used was the Pima Indians Diabetes Database~\cite{26}, an open-source dataset for diabetic diseases. Two datasets were prepared in advance: a full dataset of 392 cases from which records containing missing values were deleted, and a short version of the dataset from which 100 cases were randomly selected. The circuit configuration and associated parameters were exactly the same as those of the first experiment.

From the Pima Indians Diabetes Database, three variables were used: ‘age,’ a variable for age; ‘insulin,’ a variable for insulin concentration 2 h after the oral glucose tolerance test (OGTT); and ‘glucose,’ a variable for blood glucose concentration 2 h after the OGTT. Regarding the relationship between insulin and glucose concentrations at 2 h after the OGTT, it is not easy to determine clinically which influences which, and this was considered to be a task of clinical significance.

\begin{figure}[H]
\centering  
\includegraphics[clip,scale=0.3]{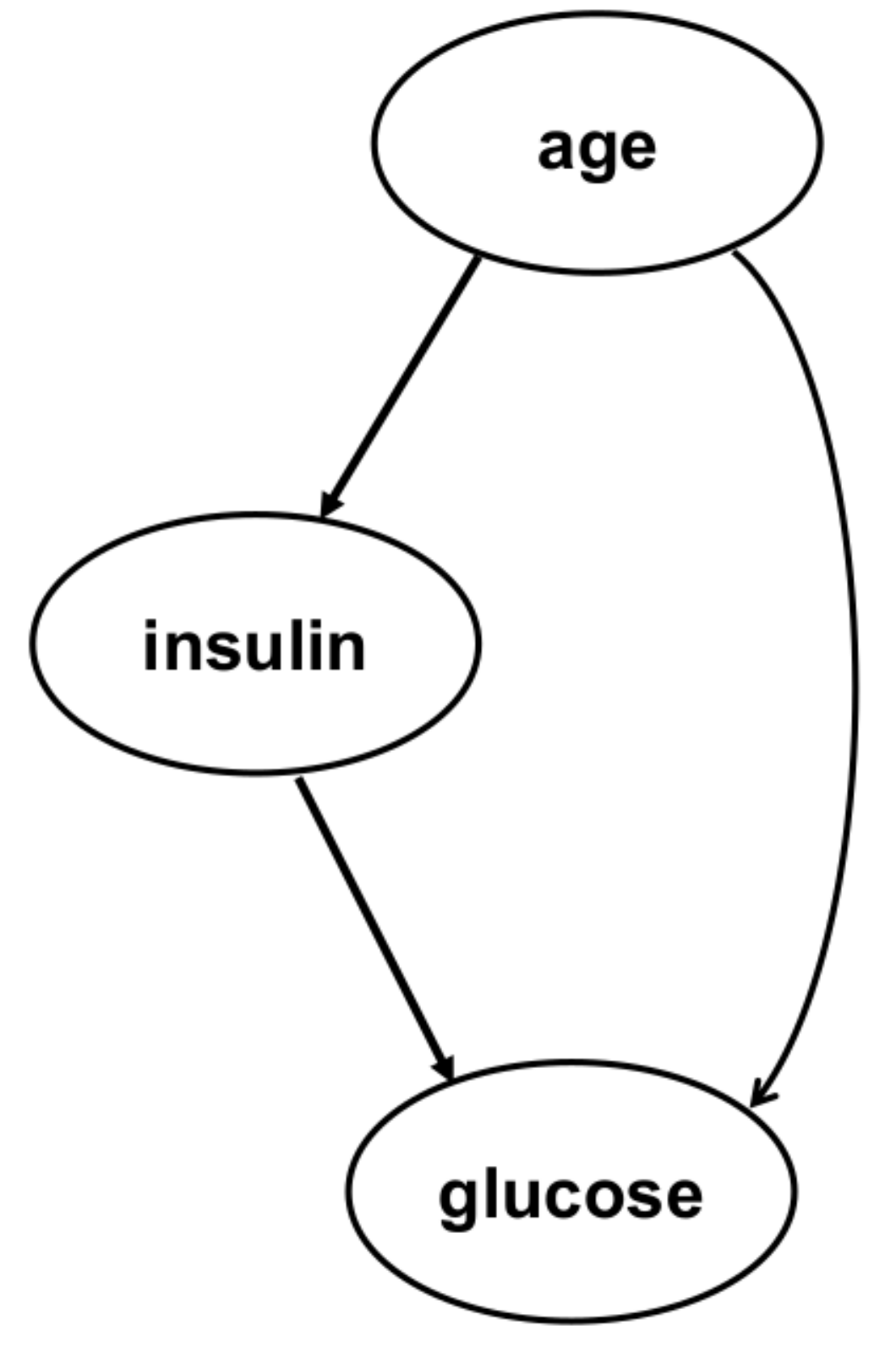}
\caption{\footnotesize\textbf{Experimental results using the Pima Indians Diabetes Database.} The ellipses represent the variables, and the arrows represent the directions of causality. ‘age’, a variable for age; ‘insulin’, a variable for insulin concentration 2 h after the oral glucose tolerance test (OGTT); and ‘glucose’, a variable for blood glucose concentration 2 h after the OGTT.}    
\label{figure6}
\end{figure}

The results of the experiment showed that both qLiNGAM and DirectLiNGAM with Gaussian kernels identified the causal relationship in Fig. \ref{figure6} for both the full example dataset (392 examples) and the short version dataset (100 randomly selected examples). As shown in Fig. \ref{figure6}, the left-hand path shows that ‘age’ affects ‘insulin,’ which, in turn, affects ‘glucose,’ and the right-hand path shows that ‘age’ directly affects ‘glucose.’ Because it was assumed that there were other factors between age and insulin concentration that were not used in this study, such as obesity, the clinically valid relationship can be confirmed, which is depicted by the right-hand path, where age affects blood glucose via a path other than insulin concentration. Thus, a case where qLiNGAM can present a causal structure even for cases that are not easy to determine clinically has been confirmed.

\subsection*{Demonstration on IBMQ}
The final experiment was to evaluate the performance of qLiNGAM using real quantum hardware. More specifically, the experiment was conducted using the $ibm\_kawasaki$ 27-qubit quantum device. Qiskit, a python library, was used to access this device. The data used was a randomly selected dataset of 100 cases from the UCI Heart Disease Data Set, which is the same as the short version of the dataset used in the experiment described above. As for the variables used, ‘age’, ‘cp’, and ‘exang’ were also used.

The circuit configuration is essentially the same as in the previous experiments, but it is difficult to obtain correct results when using existing real quantum hardware. More specifically, when the number of qubits is 5 and the number of depths is 2, it is difficult to obtain correct calculation results due to the large influence of errors. Therefore, in our experiments using real quantum hardware, a quantum circuit was configured with the number of qubits set to 4 and the number of depths set to 1. Note that even under these conditions, clinically valid causal relationships were identified in simulations using the short version of the dataset.

Among the 27 qubits in $ibm\_kawasaki$, four linearly connected qubits (specifically, q0, q1, q4, and q7) were selected to try to reduce the influence of various errors that occur in real quantum hardware, especially CNOT errors, on calculation accuracy. In addition, to mitigate the readout error of each qubit, the readout error mitigation routine in Qiskit was used to calculate the Gram matrices. In more detail, a $16\times16$ calibration matrix was prepared and applied to the obtained calculation results to correct the four qubit readout results.

When qLiNGAM was implemented with $ibm\_kawasaki$ and applied to the short version of the dataset, qLiNGAM identified the causal relationship described in Fig. 5a. Therefore, even when qLiNGAM was implemented with real quantum hardware, qLiNGAM was shown to be able to estimate valid causal structures with a smaller amount of data. 

\section*{Discussion}
\begin{figure}[H]
\centering  
\includegraphics[clip,scale=0.3]{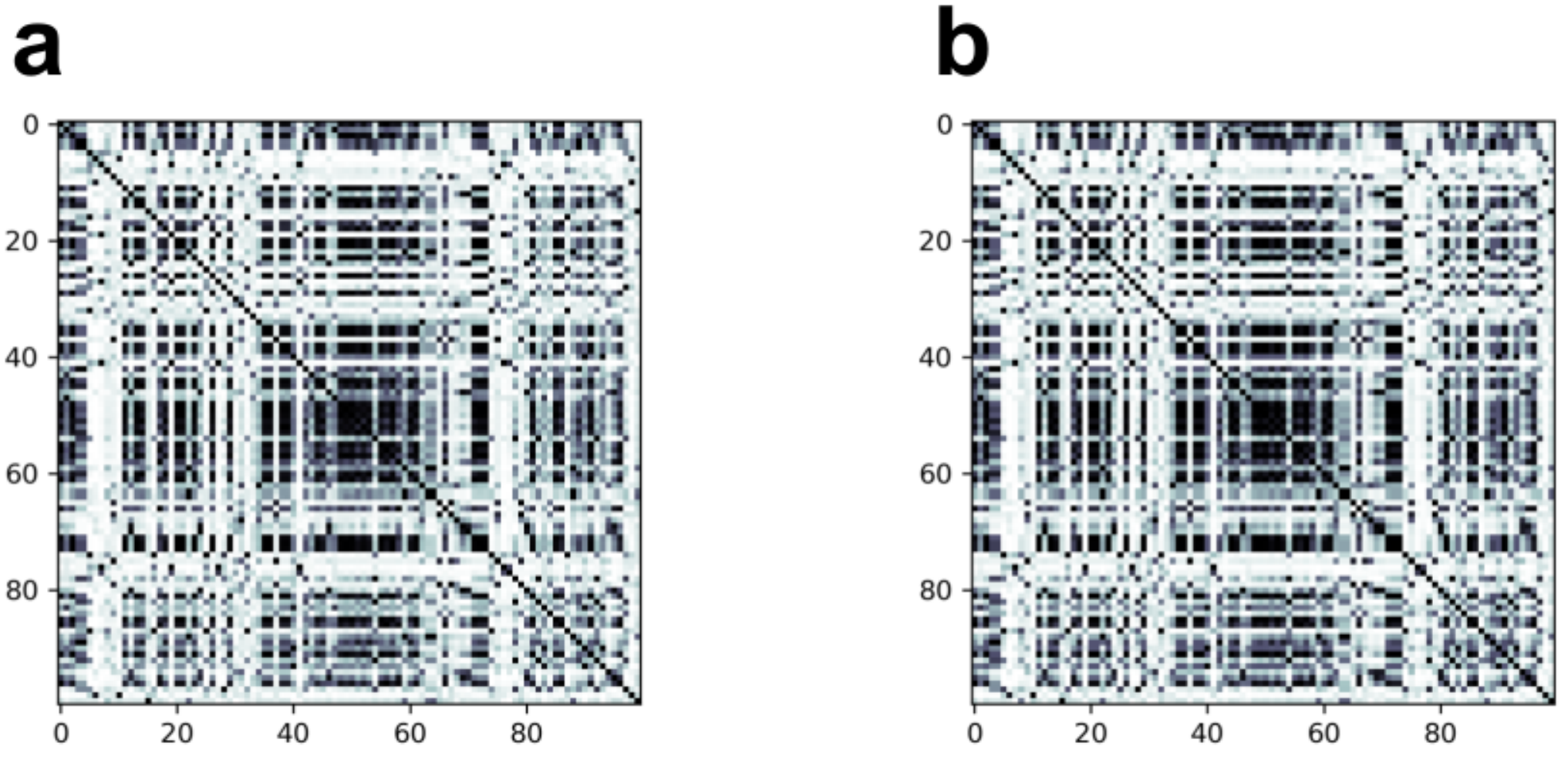}
\caption{\footnotesize\textbf{Gram matrices of the quantum kernel using the variable ‘age’ in the short version dataset with 100 cases selected from the UCI Heart Disease Data Set.} Each element of the Gram matrix is normalized from 0 to 1. The stronger the black color, the closer it is to 1. \textbf{a}, The Gram matrix using the $ibm\_kawasaki$. \textbf{b}, The theoretical values of Gram matrix of the quantum kernel.}    
\label{figure7}
\end{figure}

In this study, qLiNGAM was developed, which applies quantum kernels to DirectLiNGAM, and experiments were conducted using artificial data. qLiNGAM was able to estimate causal structures that could not be identified by DirectLiNGAM using Gaussian kernels in 14 out of 100 datasets. Thus, the use of quantum kernels suggests the possibility of estimating causal structures that cannot be detected by conventional methods. Next, qLiNGAM was applied to real-world medical data and the cases, in which clinically valid causal structures that were not easily determined at first glance could be estimated and in which clinically valid causal structures could be estimated with a smaller number of examples than those estimated by existing methods using Gaussian kernels, were confirmed. Thus, the application of qLiNGAM to real-world medical data suggests the possibility of discovering new medical knowledge and contributing to the solution of medical issues. Finally, through experiments using IBMQ, it was shown that qLiNGAM can be implemented on real quantum hardware.

qLiNGAM was successfully implemented on real quantum hardware because the quantum kernel calculated by $ibm\_kawasaki$ was close to the theoretical values, due to the adjustment of the number of depths and qubits and the error mitigation methods. With the use of the variable ‘age’ as an example, one of the variables in the UCI Heart Disease Data Set used in the previous section, the Gram matrix of quantum kernel with the short version with only 100 cases is shown in Fig. \ref{figure7}. It can be seen that the Gram matrix of the quantum kernel using the $ibm\_kawasaki$ is very close to the theoretical values. Thus, it is important to have an implementation that controls various errors of real quantum hardware and also maintains the accuracy of the algorithm.

\begin{figure}[H]
\centering  
\includegraphics[clip,scale=0.3]{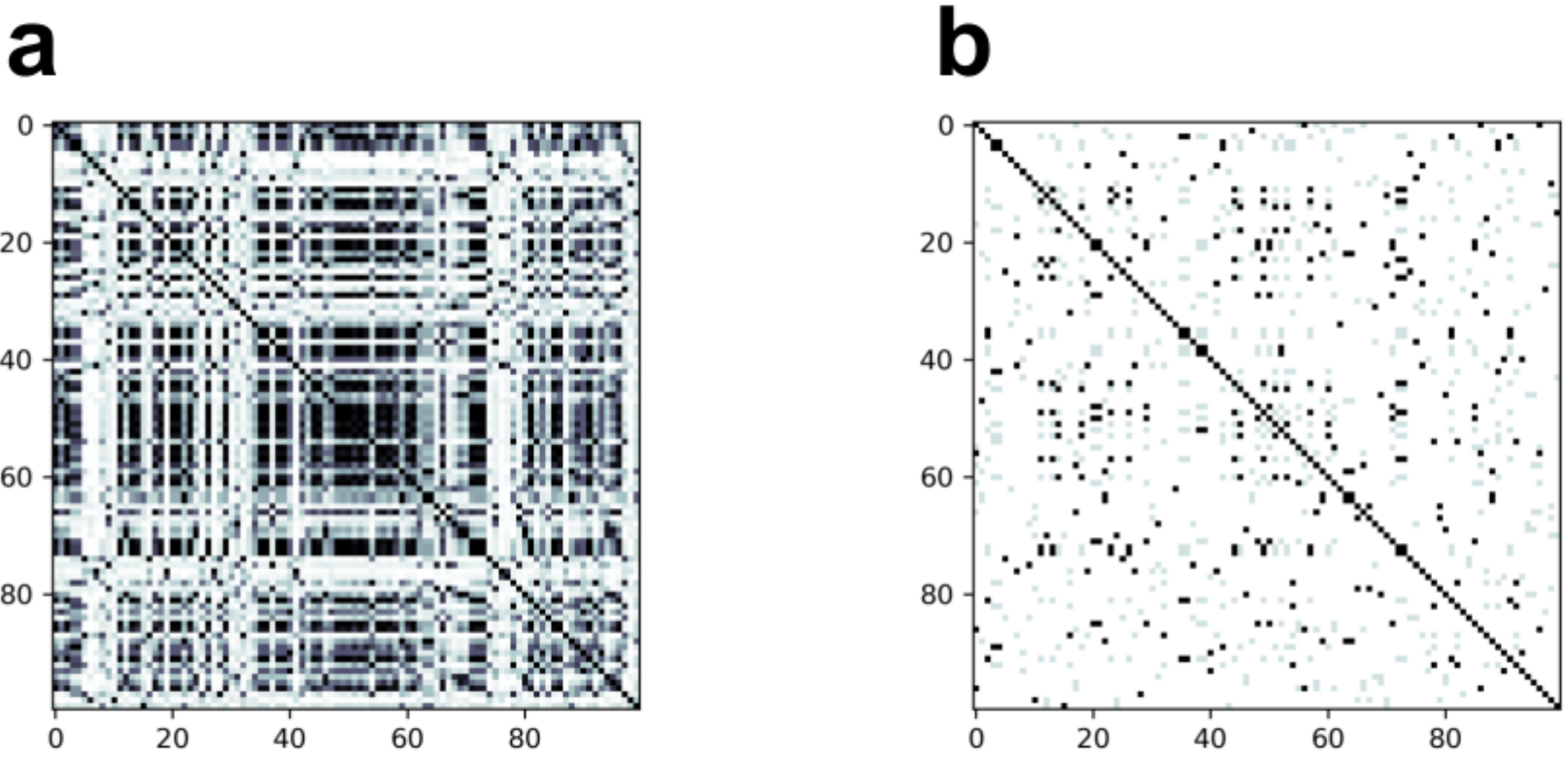}
\caption{\footnotesize\textbf{Gram matrices using the variable ‘age’ in the short version dataset with 100 cases selected from the UCI Heart Disease Data Set.} Each element of the Gram matrix is normalized from 0 to 1. The stronger the black color, the closer it is to 1. \textbf{a}, The Gram matrix of the quantum kernel. \textbf{b}, The Gram matrix of the Gaussian kernel.}    
\label{figure8}
\end{figure}

Although it is not fully clear why qLiNGAM was able to estimate the causal structure with a smaller amount of data than those used by existing algorithms using Gaussian kernels, it is possible that the Gram matrices generated using quantum calculation with an IQP circuit, which have significantly different values than those of the Gaussian kernels, allowed the evaluation of the independence with higher accuracy. Fig. \ref{figure8} shows Gram matrices of the quantum kernel and the Gaussian kernel using the variable ‘age’ in the short version dataset with 100 cases selected from the UCI Heart Disease Data Set. Overall, the Gram matrix of the quantum kernel is denser than that of the Gaussian kernel.

\begin{figure}[H]
\centering  
\includegraphics[clip,scale=0.3]{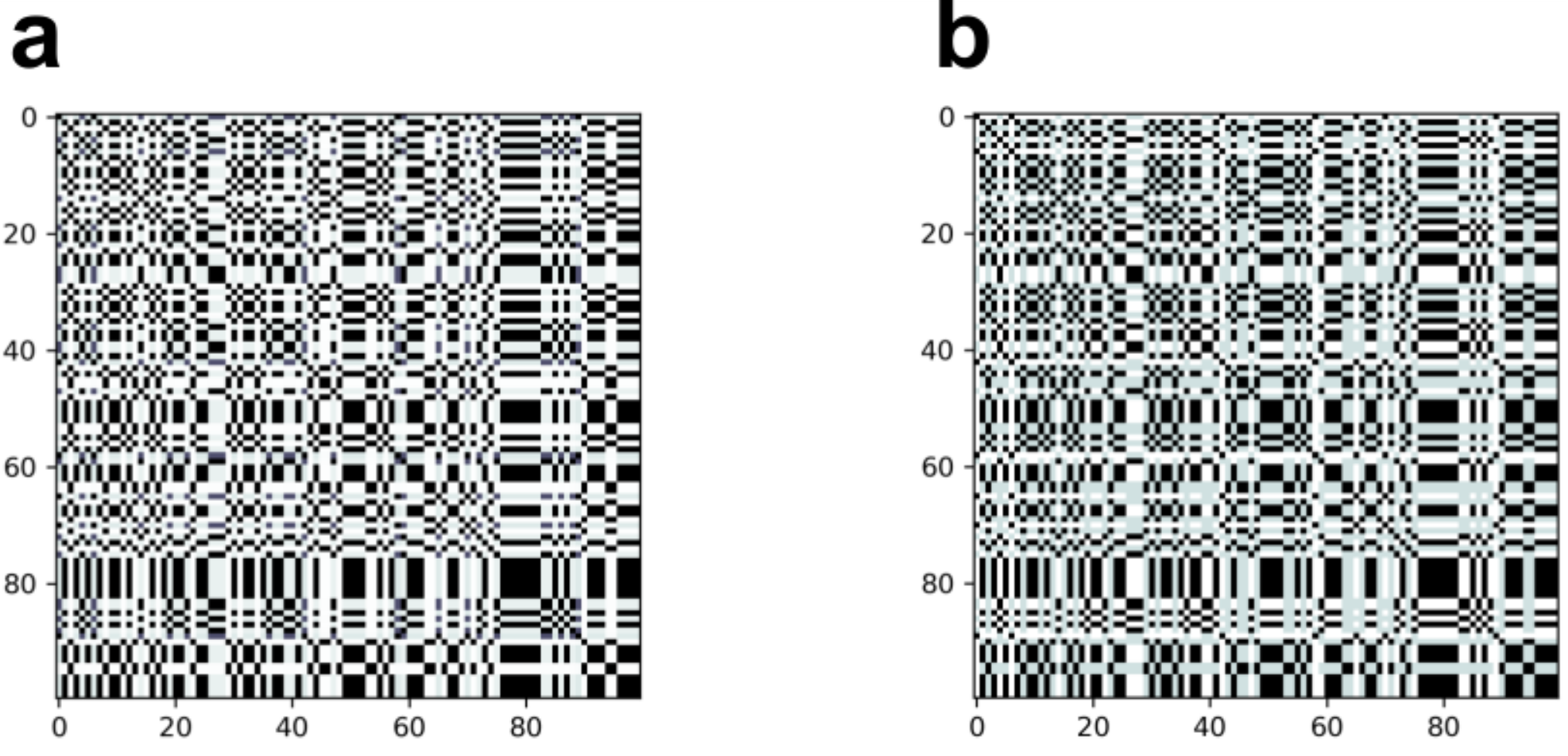}
\caption{\footnotesize\textbf{Gram matrices using the variable ‘cp’ in the short version dataset with 100 cases selected from the UCI Heart Disease Data Set.} Each element of the Gram matrix is normalized from 0 to 1. The stronger the black color, the closer it is to 1. \textbf{a}, The Gram matrix of the quantum kernel. \textbf{b}, The Gram matrix of the Gaussian kernel.}    
\label{figure9}
\end{figure}

On the other hand, with the use of the variable ‘cp’ in the UCI Heart Disease Data Set used as an example, the Gram matrix of the Gaussian kernel is denser than that of the quantum kernel, and both Gram matrices have relatively similar shapes, unlike using the variable ‘age’ (Fig. \ref{figure9}).

As shown above, the Gram matrices of the Gaussian and the quantum kernels are different depending on the type of input data and variables, and this difference may produce a different assessment of independence. In the future, a more elaborate analysis of the properties of independence measures using quantum kernels is needed, such as a quantitative evaluation of these properties through a permutation test.

In addition, qLiNGAM has several technical challenges that need to be solved in the future. For example, the quantum circuit needs to be improved to design the quantum kernel. In this study, a quantum kernel based on the IQP circuit was built; however, it is not easy to decide whether the IQP circuit is really suitable and how many depths and qubits are appropriate. In particular, it is desirable to use more qubits to create more complex kernels; however, it is known that, as the number of qubits increases, the Gram matrix of the quantum kernel becomes sparser and approaches the identity matrix, and improvement methods are being sought~\cite{27}. In addition, there is room for improvement in the selection of the feature map function and in the scaling of variables. In this study, the selection of such function was done heuristically; however, a theoretical background for the axis of search is desired in the future. Furthermore, implementation on real quantum hardware is also an important issue. It is expected that the number of qubits and the error rate will be improved in the future, and it will be necessary to adjust qLiNGAM according to the improvement of the performance of real quantum hardware.

Much of the discovery of new medical knowledge has so far been generated empirically, and the experience of clinicians should continue to be respected. By contrast, the amount of medical knowledge required for clinical practice is enormous, and it is becoming increasingly difficult to discover all new knowledge from clinicians' experience alone. It is hoped that the results of this study will support clinicians' hypothesis formation by detecting novel medical knowledge from real-world medical data. In particular, in cases where the number of samples obtained is smaller than the number of variables obtained, such as real-world medical data on rare diseases, the strength of the study algorithm, which can infer a reasonable causal relationship from a smaller amount of data, might be demonstrated. More specific applications are expected to include drug repositioning, such as the search for novel drug responses, causal search using an integrated database of genomic and clinical data, and detection of the relationship between lifestyle habits and disease outcomes in the health technology field.

\section*{Conclusions}
In this study, qLiNGAM was developed and applied to real-world medical data to verify whether clinically valid causal relationships could be identified. Furthermore, qLiNGAM was implemented on real quantum hardware in an experiment using IBMQ. With the use of a quantum kernel, there was a case where clinically valid causal relationships was identified from a smaller amount of medical data than the amount needed by DirectLiNGAM with Gaussian kernels. This suggests that the application of qLiNGAM to real-world medical data has the potential to discover new medical knowledge and can contribute to solving problems, especially when the number of samples obtained is smaller than the number of variables available.

\section*{Declarations}
\subsection*{Ethics and approval and consent to participate}
\hspace*{-1em}Not applicable
\subsection*{Competing interests}
\hspace*{-1em}The author declares that he has no competing interests.
\subsection*{Funding}
\hspace*{-1em}Not applicable
\subsection*{Acknowledgments}
\hspace*{-1em}This work was supported by the IPA MITOU Target Program.

\end{multicols}
\end{document}